\documentstyle[aps,preprint,epsf]{revtex}
\begin{document}
\title{Canonical and Microcanonical Calculations for Fermi Systems}
\author{Scott Pratt}
\address{Department of Physics and Astronomy and\\
National Superconducting
Cyclotron Laboratory,\\
Michigan State University, East Lansing, MI 48824~~USA}
\date{\today}
\maketitle

\begin{abstract}
Recursion relations are presented that allow exact calculation of canonical and
microcanonical partition functions of degenerate Fermi systems, assuming no
explicit two-body interactions. Calculations of the level density, sorted by
angular momentum, are presented for $^{56}$Ni. The issue of
treating unbound states is also addressed.
\end{abstract}


Statistical models have played a central role in the study of the nucleus since
Bohr's model of the compound nucleus in 1936\cite{bohr,weisskopf}. Level
densities and occupation probabilities are important in determining reaction
rates at finite temperature in astrophysical studies\cite{rauscher} and
heavy-ion reactions\cite{friedman}. Level densities also dominate the behavior
of low-energy neutron scattering\cite{neutronlevels}. A nearly identical set of
problems appears in atomic physics when modeling highly excited electronic
configurations in heavier atoms. There, level densities and occupation
probabilities play a pivotal role in understanding the opacity of interstellar
dust which affects the interpretation of a variety of astrophysical
measurements\cite{opacity_project,seaton}. Mesoscopic systems are also
beginning to play an increasingly important role in other fields, such as the
study of quantum dots or atomic clusters. In this letter, we present methods to
perform exact calculations of both canonical and microcanonical
quantities. These methods involve the counting of all possible arrangements of
fermions amongst independent single-particle levels. We apply these methods to
the example of counting levels in $^{56}$Ni. Even though explicit residual
interactions are ignored, the methods allow insight into a variety of problems
involving finite Fermi systems. We also discuss possible extensions of these
pictures that would treat residual interactions.

The calculations presented here are based on recursive algorithms. The
algorithms can be considered as extensions to techniques used for
multi-particle symmetrization which has been applied in the context of heavy
ion collisions for both Bose and Fermi systems\cite{prattplb,prattbauer} and
extended to weakly interacting Bose gases\cite{mekjian}. The simplest
application considers the canonical partition function, $Z_A$, for a system of
$A$ identical fermions populating a finite set of energy levels, $\epsilon_i$,
at an inverse temperature $\beta$.
\begin{equation}
Z_A(\beta)=\frac{1}{A!}\sum_{j_1\cdots j_A,
{\cal P}(j)}
\langle j_1\cdots j_A | e^{-\beta H} 
| {\cal P}(j_1\cdots j_A) \rangle.
\end{equation}
Here, the $A$ particles are treated as distinguishable. They occupy the states,
$j_1,\cdots j_A$ which are eigenstates of $H$, and ${\cal P}(j)$ refers to the
$A!$ permutations of $j_1\cdots j_A$.

The permutations may be categorized diagrammatically. As an example, a
9-particle permutation is illustrated in Fig. \ref{diagrams_fig}, where the
$n^{\rm th}$ position in the permutation is found by following the segment of
the loop that leaves from from the $n^{\rm th}$ site on the diagram. The sum
over diagrams represents the sum over permutations, and a given closed loop
represents a permutation cycle. 
\begin{equation}
C_n=\sum_ie^{-n\beta\epsilon_i}(-1)^{n-1}.
\end{equation}
All diagrams can be expressed as a product of cycles. The diagram written in
Fig. 1 may be expressed as the product, $C_1C_2^2C_4$.  The sign $(-1)^{n-1}$
arises from the number of pairwise permutations required to create the loop,
and is ignored in the case of Bosons.

One may now derive the recursion relation for the partition function by
considering summing over diagrams according to the cycle of the loop connecting
the upper left point.
\begin{eqnarray}
\label{zcanonicalrecursion_eq}
Z_A(\beta)&=&\frac{1}{A!}\left( \sum~{\rm all~diagrams~of~order~}A\right)\\
\nonumber
&=&\frac{1}{A!}\sum_n \frac{(A-1)!}{(A-n)!}
C_n (\sum~{\rm all~diagrams~of~order~}A-n)\\
\nonumber
&=&\frac{1}{A}\sum_nC_nZ_{A-n}(\beta)
\end{eqnarray}
The factor $(A-1)\cdots(A-n+1)$ in the second line of
Eq. (\ref{zcanonicalrecursion_eq}) may be understood by counting the number of
ways in which a loop of size $n$ may be connected to one point.

Eq. (\ref{zcanonicalrecursion_eq}) allows practically instantaneous calculation
of the partition function given the single-particle energy levels $\epsilon_i$.
This approach also leads to a simple expression for two-point functions,
\begin{equation}
\label{phasespace_eq}
\langle a^\dagger_ia_j\rangle = \delta_{ij}
\frac{1}{Z_A(\beta)}\sum_n e^{-n\beta\epsilon_i}(-1)^{n-1}Z_{A-n}(\beta).
\end{equation}
The four-point function might also be of interest in calculations of
residual interactions.
\begin{equation}
\label{fourpoint_canonical_eq}
\langle a^\dagger_ia^\dagger_j a_ka_l\rangle=\frac{1}{Z_A(\beta)}
\left(\delta_{il}\delta_{jk}-\delta_{ik}\delta_{jl}\right)
\sum_{n_i,n_j}e^{-n_i\beta\epsilon_i-n_j\beta\epsilon_j}
(-1)^{n_i+n_j}
Z_{A-n_i-n_j}(\beta)
\end{equation}
It is straight-forward to extend these calculations to arbitrary $n$-point
functions.

The method is easily extended to include conserved quantities. For instance,
conservation of the $z$-component of angular momentum $J_z$ can be accommodated
by summing over only those diagrams which yield a fixed sum, $M=\sum_i n_im_i$.
\begin{equation}
Z_{A,M}(\beta)=\frac{1}{A}\sum_{n,i}e^{-n\beta\epsilon_i}(-1)^{n-1}
Z_{A-n,M-nm_i}(\beta)
\end{equation}
In a similar fashion one may incorporate any other conservation laws, such as
parity, by adding an additional subscript to $Z$.  

By setting $\beta$ to zero, one may associate $Z$ with a counting of all states
regardless of energy. By setting $\beta=0$, and assigning a discretized energy
to each state, one may count the net number of states with total energy $E$.
\begin{equation}
\label{AMErecursion_eq}
N_{A,E,M}=\frac{1}{A}\sum_{n,i}(-1)^{n-1}N_{A-n,E-n\epsilon_i,M-nm_i}.
\end{equation}
The above relation requires that energy be measured in discrete units. These
units can be arbitrarily small to make the calculation approach the continuum
limit. The quantity $N$ represents the microcanonical partition function, and
when divided by the energy bin size, yields the density of states.

One may also calculate the microcanonical equivalent of the two and four-point
functions.
\begin{eqnarray}
\langle a^\dagger _ia_j\rangle&=&\delta_{ij}
\frac{\sum_n (-1)^{n-1} N_{A-n,E^*-n\epsilon_i,M-nm_i}}{N_{A,E^*,M}}\\
\langle a^\dagger_ia^\dagger_ja_ka_l\rangle &=&
\left( \delta_{il}\delta_{jk}-\delta_{ik}\delta_{jl}\right)
\frac{\sum_{n_in_j}(-1)^{n_i+n_j}
N_{A-n_i-n_j,E^*-n_i\epsilon_i-n_j\epsilon_j,M-n_im_i-n_jm_j}}{N_{A,E^*,M}}
\label{fourpoint_micro_eq}
\end{eqnarray}

Calculations proceed more quickly when particles and holes are considered
independently, then convoluted to find the number of states at a fixed $E^*$
and $M$. Considering only states with energies above the Fermi level, one first
finds $N^p_{a,E^*,M}$, the number of ways to arrange $a$ particles above the
Fermi surface such that they sum to energy $E^*$ and angular momentum
projection $M$.  Then, considering only states below the Fermi surface, one
performs the analogous calculation for $a$ holes, $N^h_{a,E^*,M}$. The number
of ways to arrange particles and holes subject to conservation of particle
number, energy and angular momentum is
\begin{equation}
\label{phole_eq}
N_{E^*,M}=\sum_{a,E^\prime,M^\prime}
N^p_{a,E^\prime-E_f,M-M^\prime}N^h_{a,E^*-(E^\prime-E_f),M^{\prime}}.
\end{equation}

Instead of counting states with a specific projection $M=\langle J_z\rangle$,
one may count $\tilde{N}_J$, the number of multiplets of a given $J$, by
converting the recursion relations for $N_M$ into a relation for
$\tilde{N}_J$.
\begin{eqnarray}
\label{JtoM_eq}
\tilde{N}_J&=&N_{J}-N_{J+1}\\
\nonumber
N_M&=&\sum_{J\ge |M|}\tilde{N}_J
\end{eqnarray}
By manipulating Eq. (\ref{AMErecursion_eq}) and Eq.s (\ref{JtoM_eq}), one may
derive a recursion relation for $\tilde{N}_J$.
\begin{equation}
\label{AJErecursion_eq}
\tilde{N}_{A,J,E}=\sum_{n,i}(-1)^{n-1}\sum_{-j_i\le m_i\le j_i}
\left( \tilde{N}_{A-n,J-nm_i,E-n\epsilon_i}\Theta(J-nm_i)
-\tilde{N}_{A-n,nm_i-J-1,E-n\epsilon_i}\Theta(nm_i-J-1) \right),
\end{equation}
where the sum over $i$ represents a sum over multiplets of a given $j$ rather
than a sum over individual states.

The simplest example to consider is that of equally-spaced non-degenerate
levels which was addressed by Ericson\cite{ericson}. This example can be
related to the problem of finding the number of ways an integer can be written
as a sum of smaller integers without using the same integer twice. The
resulting density of states which is displayed in Fig. (\ref{rho_nickel_fig})
is well matched by the Bethe formula\cite{bethe,bohrmottleson},
\begin{equation}
\rho(E)\approx\frac{\exp{\left( 2\pi\sqrt{gE^*/6}\right)}}{\sqrt{48}E^*},
\end{equation}
where $g$ is the single-particle level density and $E^*$ is the excitation
energy above the ground state. Fig. \ref{phasespace_fig} displays the two-point
function, $f_i=\langle a^\dagger_ia_i\rangle$ at a fixed excitation energy of
20 $g^{-1}$. The resulting phase space filling factor is fairly well described
by a Fermi function, represented by a solid line in Fig. \ref{phasespace_fig},
with the temperature set at $T=\sqrt{\frac{6E^*}{\pi^2g}}$.

We now consider a more realistic example where single-particle levels for
$^{56}$Ni were generated from a Hartree-Fock calculation\cite{alexhf}. An
infinite external potential was added to confine the particles to within a 10
fm sphere, transforming the continuum into a set of densely packed discrete
states. All levels within 50 MeV of the Fermi surface were included. The lowest
20 single-particle energy levels are listed in Table \ref{hflevels_tab}.

\begin{table}
\caption{Single-particle energy levels of $^{56}$Ni from Hartree-Fock
calculations in MeV.\label{hflevels_tab}}
\begin{tabular}{|c|d|d||c|d|d|}
level&Proton&Neutron&level&Proton&Neutron\\ \hline
 $s_{1/2}$ & -32.59 & -42.61 & $g_{9/2}$ &  3.93 & -5.33 \\
 $p_{3/2}$ & -24.23 & -34.10 & $d_{5/2}$ &  8.17 & -0.53 \\
 $p_{1/2}$ & -22.56 & -32.48 & $d_{3/2}$ &  9.36 &  0.94 \\
 $d_{5/2}$ & -15.33 & -25.04 & $s_{1/2}$ &  9.41 &  1.32 \\
 $s_{1/2}$ & -11.46 & -21.28 & $g_{7/2}$ & 12.11 &  3.30 \\
 $d_{3/2}$ & -11.10 & -20.87 & $h_{11/2}$ & 14.37 &  5.52 \\
 $f_{7/2}$ & -6.01 & -15.53 & $p_{3/2}$ & 21.42 &  7.68 \\
 $p_{3/2}$ & -1.90 & -11.41 & $p_{1/2}$ & 14.53 &  8.16 \\
 $p_{1/2}$ & -0.52 & -9.92 & $f_{7/2}$ & 14.95 &  8.66 \\
 $f_{5/2}$ &  0.80 & -8.67 & $f_{5/2}$ & 15.92 & 10.00
\end{tabular}
\end{table}

The Fermi energy was set at -15.5 MeV for neutrons and -6.0 MeV for
protons. The energy was discretized in units of 0.25 MeV. One may then couple
the resulting level density for neutrons with the proton level density to
obtain the level density of the composite system.

Figure \ref{rho_nickel_fig} displays the resulting density of states for three
cases. First, the density of states is displayed for the case where all
single-particle levels are included, even those that are unbound. Secondly,
results are shown for a calculation where only the $fp$ and $g_{9/2}$ shells
are considered, as was implemented in a Monte Carlo shell-model calculation of
Nakada and Alhassid\cite{alhassid}. Ormand\cite{ormand},
Langanke\cite{langanke} and Dean\cite{dean} have performed similar studies.

Continuum levels should not be included in the treatment\cite{shlomo}.
Therefore, the recursive sums were modified to subtract the contribution from
the continuum in the following manner,
\begin{equation}
N_{A,E}=\frac{1}{A}\sum_n(-1)^{n-1}
\left(
\sum_i N_{A-n,E-\epsilon_i} -\sum_{i^\prime}N_{A-n,E-\epsilon_{i^\prime}}
\right),
\end{equation}
where the primed sums are over eigenstates of the infinite confining spherical
well without interactions. In the limit that the radius of the confining well
approaches infinity, the correction should be exact. Figure
\ref{rho_nickel_fig} displays the corrected level density for the $^{56}$Ni
example. The correction only becomes substantial above excitations of 25
MeV. At lower energies the aforementioned subset of levels seems sufficient to
reproduce the density of states.

Figure \ref{rhovsj_fig} displays the number of multiplets within specified
energy windows for $^{56}$Ni as a function of $J$. Describing the coupling of
angular momentum as a diffusive random walk\cite{ericson}, one
expects the $J$ dependence to behave as
\begin{equation}
\label{jdependenceform_eq}
N_{E,J}\propto (J+\frac{1}{2})\exp{-\beta J(J+1)},
\end{equation}
where $\beta$ is related to the number of particles and the average $j$ of the
single-particle levels used to form the angular momentum\cite{ericson}.  As
seen in Fig. \ref{rhovsj_fig} where $\beta$ was chosen to best fit the results,
the functional form is accurate except at high $J$.

The recursion relations presented here appear quite powerful in calculating
statistical quantities of excited nuclei with minimal calculational time. The
method may easily accommodate an arbitrarily large number of single-particle
levels. The $^{56}$Ni calculations required approximately 5 minutes of CPU time
on a low-end workstation, and most of that time was spent performing the
coupling of particles to holes described in Eq. (\ref{phole_eq}).  Enumeration
of states and their angular momentum is required for nuclear shell-model
calculations. In current shell-model codes, this is accomplished by
combinatorially considering all ways in which $n$ particles can be placed into
$m$ states. These methods are sufficient as long as a small number of
single-particle levels are involved. For large nuclei or as one enters or
approaches the continuum, the number of single-particle levels becomes large
and the benefits of the recursive method are more apparent. Since the recursive
method yields partition functions, it is straight-forward to extend it to the
calculation of any statistical quantity.

Although the relations are flexible in handling any model given the
single-particle levels, two-body (residual) interactions are explicitly
ignored. However, they can be included perturbatively. For example, four-point
terms can be calculated using Eq.s (\ref{fourpoint_canonical_eq}) and
(\ref{fourpoint_micro_eq}). The four-point terms can then be used to calculate
the contribution of the residual interaction in lowest order. It is our hope
that the recursive techniques presented here might be expanded to include a
framework for calculating successively higher order perturbations.

\acknowledgements{The author wishes to thank Vladimir Zelevinsky, Alex Brown
and Declan Mulhall for sharing their expertise. This work was support by the
National Science Foundation, PHY-96-05207.

\begin{figure}
\epsfxsize=0.2\textwidth \centerline{\epsfbox{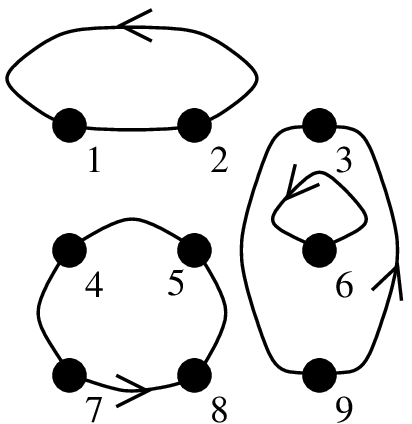}}
\caption{A sample permutation is represented by the diagram above. The
permutation of $j_1\cdots j_9$ in this case is
$j_2,j_1,j_9,j_7,j_4,j_6,j_8,j_5,j_3$ as the segment leaving the dot labeled
$n$, is connected to $j_n^\prime$ in the permutation. This diagram can be
expressed in terms of cycles by $C_1C_2^2C_4$.\label{diagrams_fig}}
\end{figure} 

\begin{figure}
\epsfxsize=0.4\textwidth \centerline{\epsfbox{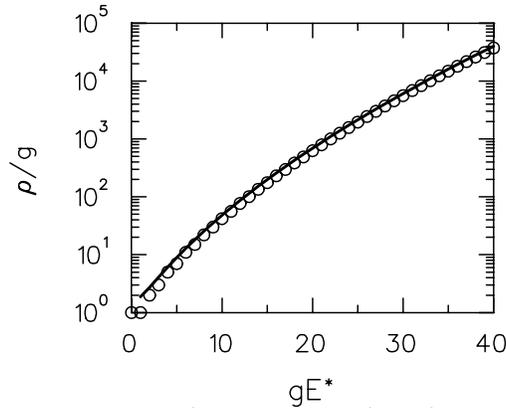}}
\caption{The level density is displayed for the case of uniformly spaced
levels. It is well approximated by the Bethe formula when the excitation energy
is much greater than the inverse single-particle level spacing,
$g^{-1}$.\label{rho_uniform_fig}}
\end{figure}

\begin{figure}
\epsfxsize=0.4\textwidth \centerline{\epsfbox{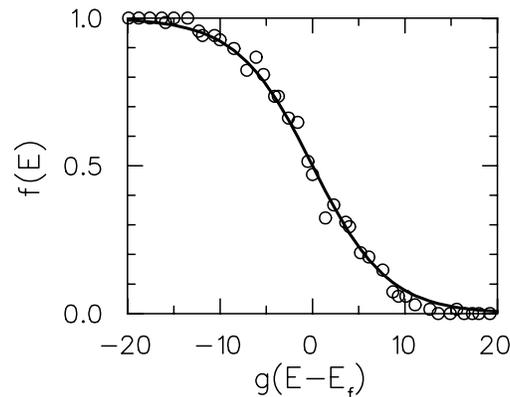}}
\caption{The occupation probability is shown for the case of uniformly-spaced
levels where the energy is fixed at 10 times the single-particle level
spacing.\label{phasespace_fig}}
\end{figure}

\begin{figure}
\epsfxsize=0.45\textwidth \centerline{\epsfbox{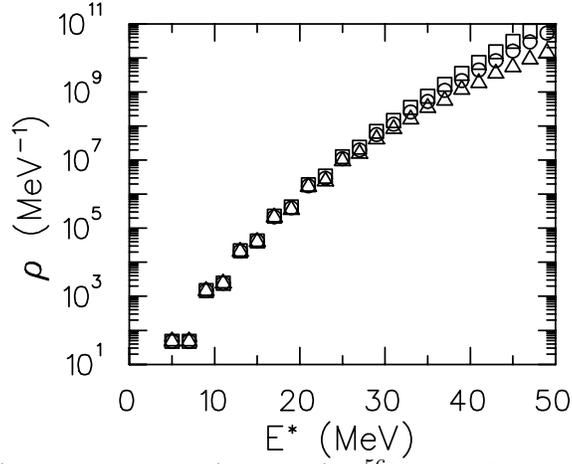}}
\caption{Calculations for the density of states for $^{56}$Ni are displayed for
three sets of states. Results which consider only $fp$ and the $g_{9/2}$ shell
are represented by triangles, including all states is represented by squares,
while subtracting continuum corrections yields the results represented by
circles.\label{rho_nickel_fig}}
\end{figure}

\begin{figure}
\epsfxsize=0.4\textwidth \centerline{\epsfbox{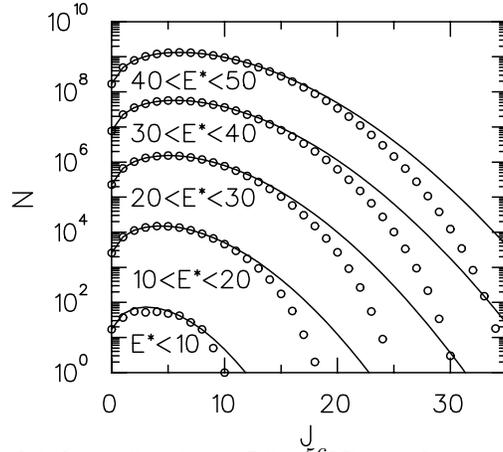}}
\caption{The number of multiplets of a given $J$ for $^{56}$Ni are shown for
the energy windows labeled above. The fits to the form shown in
Eq. (\ref{jdependenceform_eq}) are shown with lines. The form well describes
the result, except for high $J$.\label{rhovsj_fig}}
\end{figure}

\end{document}